\documentclass[letterpaper,10pt]{article} 

\usepackage{opticameet3} 

\newcommand\authormark[1]{\textsuperscript{#1}}

\usepackage{amsmath,amssymb}
\usepackage[colorlinks=true,bookmarks=false,citecolor=blue,urlcolor=blue]{hyperref} 
\usepackage{cite}
\usepackage[labelformat=simple]{subfig}
\usepackage{float}
\usepackage{caption}
\usepackage{cleveref}
\usepackage{amsmath,amssymb}
\usepackage[version=3]{mhchem}
\usepackage[per-mode=symbol]{siunitx}
\DeclareSIUnit\angstrom{\text {Å}}

\usepackage{acronym}
\acrodef{EBL}{electron beam lithography}  
\acrodef{t-SPL}{thermal scanning probe lithography}
\acrodef{EOT}{extraordinary optical transmission}
\acrodef{PMMA/MA}{polymethyl (methacrylate-co-methacrylic acid)} 
\acrodef{PPA}{phthalaldehyde polymer}
\acrodef{AFM}{atomic force microscopy}
\acrodef{IR}{infrared}

\begin{document}

\title{Thermal Scanning Probe Lithography for Fabrication of Perforated Metallic Films}

\copyrightyear{2025}

\author{Paloma E. S. Pellegrini,\authormark{1,*} Francisco T. Orlandini,\authormark{1} Silvia V. G. Nista,\authormark{1} Stéphane Lanteri,\authormark{2} Hugo Enrique Hernández-Figueroa,\authormark{3} and Stanislav Moshkalev \authormark{1}}

\address{\authormark{1} 
Center for Semiconductor Components and Nanotechnology, Universidade Estadual de Campinas, Brazil\\
\authormark{2}Université Côte d'Azur, Inria, CNRS, LJAD, France\\
\authormark{3}
School of Electrical and Computer Engineering, Universidade Estadual de Campinas, Brazil}

\email{\authormark{*}palomap@unicamp.br} 

\begin{abstract} 
Thermal scanning probe lithography offers high resolution and versatility, making it a promising alternative for fabricating photonic devices. Here, we introduce a new method that expands its applications by enabling direct fabrication of arbitrary perforated patterns on a silver film.
\end{abstract}

\vspace{1cm}

The propagation of electromagnetic waves through perforated metallic films, at a subwavelength scale, leads to strong transmissive resonances, referred to \ac{EOT}~\cite{huang2011extraordinary}. It has led to to various photonic applications such as sensing~\cite{okamoto2024design}, color filtering and perfect absorbers~\cite{yan2020perfect}. As the simulation of nanophotonic devices advances~\cite{ so2020deep}, we see an increasing demand for fabricating complex structures with high resolution~\cite{okamoto2024design}. Therefore, it is essential to expand and propose new fabrication techniques that can account for the development of photonics devices.




One the most established fabrication techniques is the \ac{EBL}. However, the setup for an \ac{EBL} is not simple. It requires high-vacuum, cleanrooms, and a complex electron-optics apparatus. Moreover, this technique has the intrinsic drawback of proximity effects, that can impair patterning~\cite{chen2015nanofabrication}.



In this context, \ac{t-SPL} offers an advantageous fabrication alternative~\cite{howell2020thermal}. It is a top-down technique that relies on a nanometric scanning probe that heats controllably. When it is in contact with a thermal sensitive resist-coated-sample patterning occurs directly, also achieving sub-10nm resolution~\cite{ryu2017sub}. It overcomes the obstacle of proximity effect from \ac{EBL} and it operates in standard room conditions. The fabrication of different photonic devices such as metasurfaces~\cite{zhang2023fabrication} has, recently, been developed with \ac{t-SPL}. So far, to the best of our knowledge, \ac{t-SPL} has only been applied similarly to lithography with a positive-tone resist. Meaning, those where the pattern is transferred to the resist and the material is further deposited and adhered on the patterned regions. Hence, it limits fabrication of perforated metasurfaces. 

\begin{figure}
    \centering
    \subfloat[]
  {\label{fig:transm}\includegraphics[width=0.22\textwidth]{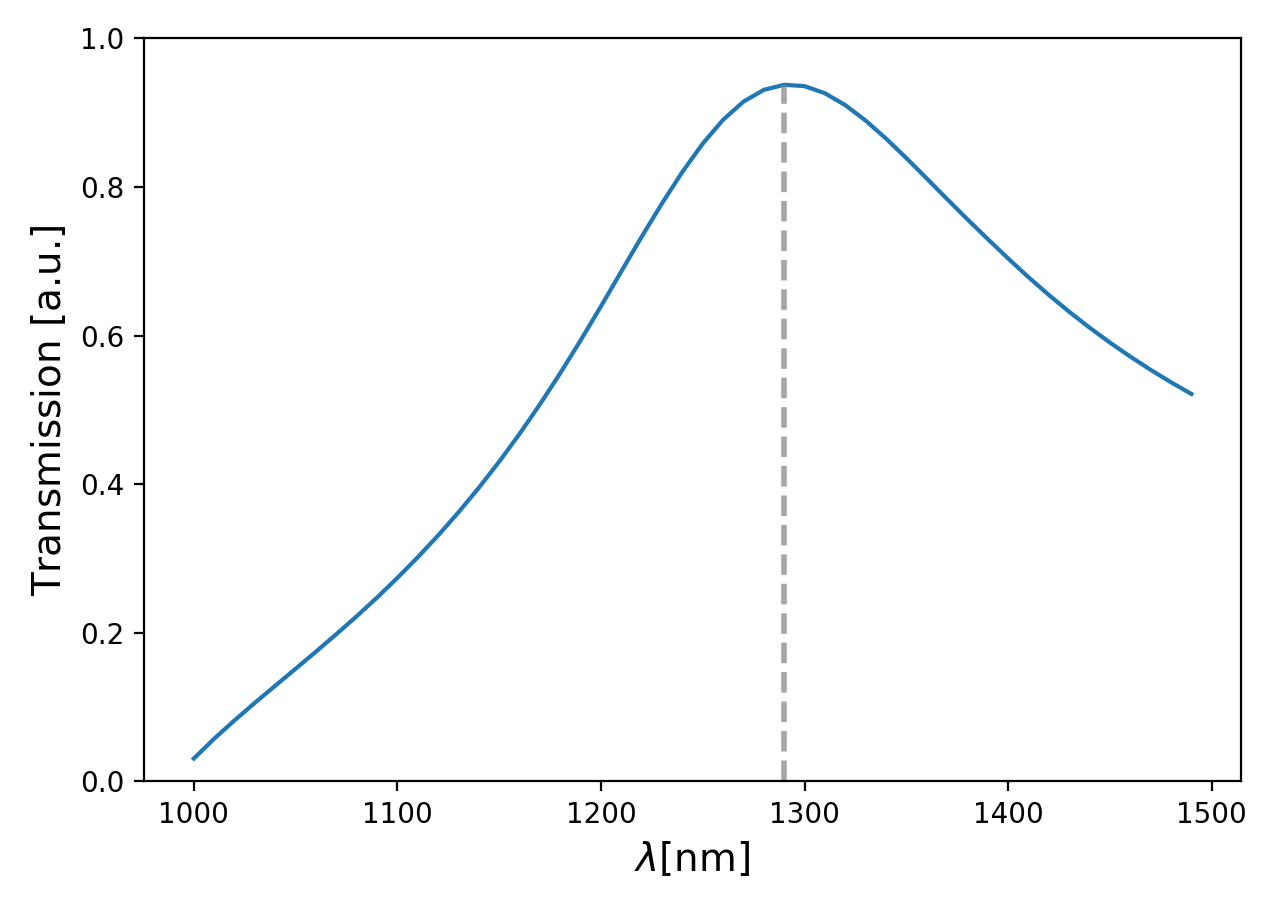}}
  \qquad
    \subfloat[]{\label{fig:esquema}\includegraphics[width=0.14\textwidth]{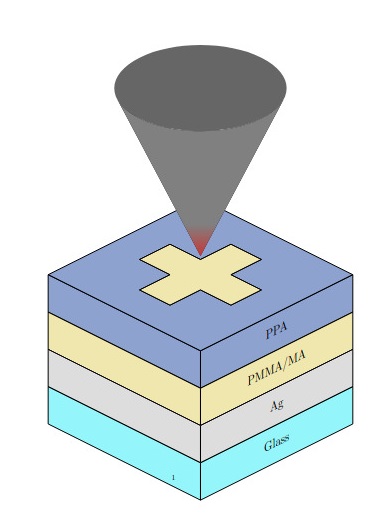}}
    \qquad
    \subfloat[]{\label{fig:afm}\includegraphics[width=0.16\textwidth]{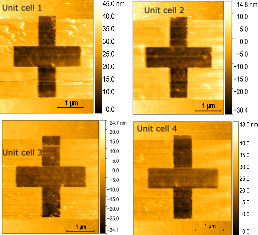}}
   \qquad
    \subfloat[]{\label{fig:final}\includegraphics[width=0.16\textwidth]{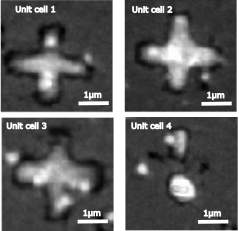}}
   \caption{Illustration of \ac{t-SPL}, (a). \ac{AFM} images of patterns transfer, (b). Perforated unit cells on the \ce{Ag} film, (c).}
    \label{fig:Results}
\end{figure}

Here, we propose a methodology using \ac{t-SPL} to fabricate perforated patterns on a silver thin film, without any specific negative-tone resists. 
It was validated with micrometric patterns inspired by the cross-shaped geometries found in ~\cite{yan2020perfect}. Initial numerical experiments have shown that such geometries have potential
applicability as filters, as shown in~\cref{fig:transm}, for the \ac{IR} region of the spectrum.
The calculations were performed using the Finite Element Method with $\mathrm{H}(\textrm{curl},\Omega)$-conforming elements, and the precise details of the chosen numerical scheme are the subject of ongoing work.


The proposed methodology follows the standard bilayer liftoff configuration of \ac{t-SPL}~\cite{pellegrini2024assembly}, while also integrating a wet etch of the metallic film. It can be summarized in five main sequential steps, detailed throughout this work: preparation of samples, deposition of metal and resists, lithography, etching, and liftoff.

The silica substrates were cleaned using sonication in acetone and isopropyl alcohol, followed by low-pressure \ce{O2} plasma treatment (Tergeo Plasma Cleaner, Pie Scientific) to remove residual contaminants. 

A 30 nm silver film was deposited on the clean substrates via electron beam evaporation, at a base pressure of $1.7 \times 10^{-7}$ Torr and \SI{115}{\milli \ampere}, at \SI{2.5}{\angstrom / \second}. Then, the resists for \ac{t-SPL} were spin-coated on the metal film, each for \SI{30}{\second}. On top of the silver film, we deposited \SI{30}{nm} of \ac{PMMA/MA}, the liftoff resist in a 1:3 solution, at 4000 rpm for \SI{30}{\second} and soft-baked for \SI{90}{\second} at \SI{180}{\degree C}. For the top layer, we used a thermal resist that sublimates when in contact with the hot probe of the \ac{t-SPL}. The thermal resist used was \ac{PPA}, and it was coated at 6000 rpm and then baked at \SI{110}{\degree C} for \SI{120}{\second}, resulting in a thickness of \SI{20}{nm}. We emphasize that this is the minimum depth that should be sublimated during \ac{t-SPL}.

Once the samples were properly coated, they were ready for patterning through \ac{t-SPL}. The apparatus used was the NanoFrazor, from Heidelberg Instruments. Its nanometric probe is integrated with an \ac{AFM} system, allowing \emph{in-situ} topographic evaluation of the pattern transfer~\cite{howell2020thermal}. This reading is also performed before patterning, so the surface can be mapped and a leveling correction made during writing. The probe temperature was set to \SI{960}{\degree C}, and it approached the surface of the sample according to the predetermined pattern, as illustrated in \cref{fig:esquema}. This temperature was optimized in previous experiments to properly etch the \ac{PPA}. In our case, the design of cross-shaped unit cells was \SI{2.5}{\mu m} long and \SI{645}{\nm} wide.

Patterning was successfully achieved via \ac{t-SPL}, as it can be seen in \cref{fig:afm}. In average, the transferred patterns exhibited crosses of \SI{2.48}{\mu m} of length and \SI{638}{nm}. Therefore, a high resolution of sub-10nm was accomplished. Although, when mapping the surface, we noticed some oxidation spots. These marks can make the surface of the samples highly uneven and, hence, the \ac{t-SPL} apparatus has to make significant corrections to enable writing. We can see the influence of these corrections on Unit Cell 3, where the right arm of the cross is slightly blurred. These issues also led to different sublimation processes of the resist, and therefore, different etching depths of the \ac{PPA}. The depth of the pattern of Unit cell 4 is only \SI{12}{nm}, while the others are \SI{20}{nm}, as expected. For this reason, we expect that Unit Cell 4 is compromised. 

After patterning, the samples underwent an immersion in ethyl alcohol, for \SI{40}{\second} to etch the \ac{PMMA/MA} layer that was exposed after the lithography. At this step, we let the reader know that \ac{PPA} is tolerant to ethyl alcohol while \ac{PMMA/MA} is not. Hence, \ac{PPA} acts as a mask for the unpatterned regions of the samples. The samples were rinsed with isopropyl alcohol and \ce{N2} dried. Finally, the samples were immersed in a 0.05 M aqueous solution of \ce{Fe(NO3)3}. Both resists used are tolerant to iron(III) nitrate, although this salt etches \ce{Ag}. Thus, the \ce{Ag} film exposed in the cross patterns was eliminated under 2 minutes, leaving only the silica glass of the substrates.

We observe the final results in \cref{fig:final}. As anticipated, Unit Cell 4 was not perforated due to issues with surface leveling and patterning depth. The oxidation marks in Unit Cell 3 led to a less defined shape. As for the other cells, where the surface was flatter and showed less oxidation, the final perforated shapes matched the desired design.

While the deposition and storage of the \ce{Ag} films still have room for improvement to avoid oxidation~\cite{wang2017strong}, we proposed a novel methodology using \ac{t-SPL} to fabricate perforated metasurfaces. Working similarly to lithography with a negative-tone resist, \ac{t-SPL} has proven to be efficient in a simpler setup, enabling \emph{in-situ} evaluation of the patterns while still achieving sub-10 nm pattern transfer. Moreover, it was not necessary to use any negative-tone resists; the material used was the same as for standard lithography. As far as we are aware, we have introduced a novel integration of \ac{t-SPL} with perforated metallic patterns. This advancement expands the range of solid fabrication techniques available for achieving complex photonic designs.

\footnotesize{

\section*{Acknowledgments}

This research was funded by the Brazilian Council for Scientific and Technological Development (CNPq), projects: 408476/2022-2, 406193/2022-3, and 314539/2023-9 (HEHF), and by the São Paulo Research Foundation (FAPESP), projects 2021/11380–5 (CPTEn), 2021/00199–8 (SMARTNESS), 2022/11596-0 (EMU), and 2024/04835-4 (INRIA Sophia-Antipolis). The authors used the facilities of the Brazilian Nanotechnology National Laboratory (LNNano) at the Centre for Research in Energy and Materials (CNPEM), proposal MNF-20250217.

\bibliographystyle{opticajnl}
\bibliography{sample_modified}

\begin{thebibliography}{10}
\newcommand{\enquote}[1]{``#1''}

\bibitem{huang2011extraordinary}
S.~Huang \emph{et~al.}, {\protect\JournalTitle{Applied Physics Letters}}
  \textbf{98} (2011).

\bibitem{okamoto2024design}
K.~Okamoto \emph{et~al.}, in \emph{Photonics,}  vol.~11 (MDPI, 2024), p. 292.

\bibitem{yan2020perfect}
Z.~Yan \emph{et~al.}, {\protect\JournalTitle{Nanomaterials}} \textbf{11}, 63
  (2020).

\bibitem{so2020deep}
S.~So \emph{et~al.}, {\protect\JournalTitle{Nanophotonics}} \textbf{9},
  1041--1057 (2020).

\bibitem{chen2015nanofabrication}
Y.~Chen, {\protect\JournalTitle{Microelectronic Engineering}} \textbf{135},
  57--72 (2015).

\bibitem{howell2020thermal}
S.~Howell \emph{et~al.}, {\protect\JournalTitle{Microsystems \&
  nanoengineering}} \textbf{6}, 21 (2020).

\bibitem{ryu2017sub}
R.~C. \emph{et~al.}, {\protect\JournalTitle{ACS nano}} \textbf{11},
  11890--11897 (2017).

\bibitem{zhang2023fabrication}
W.~Zhang \emph{et~al.}, {\protect\JournalTitle{ACS Applied Materials \&
  Interfaces}} \textbf{15}, 13517--13525 (2023).

\bibitem{pellegrini2024assembly}
P.~Pellegrini \emph{et~al.}, {\protect\JournalTitle{Nanomanufacturing}}
  \textbf{4}, 173--186 (2024).

\bibitem{wang2017strong}
X.~Wang \emph{et~al.}, {\protect\JournalTitle{small}} \textbf{13}, 1700044
  (2017).

\end{thebibliography}
}


\end{document}